\documentclass[longauth]{aa} % for the long lists of affiliations 
\usepackage{natbib}
\bibpunct{(}{)}{;}{a}{}{,} % to follow the A&A style

\usepackage{graphicx}
\usepackage{txfonts}
\begin{document}
\title{High zenith angle observations of \mbox{PKS\,2155-304} with the MAGIC-I telescope}

\author{
 J.~Aleksi\'c\inst{1} \and
 E.~A.~Alvarez\inst{2} \and
 L.~A.~Antonelli\inst{3} \and
 P.~Antoranz\inst{4} \and
 M.~Asensio\inst{2} \and
 M.~Backes\inst{5} \and
 U.~Barres de Almeida\inst{6} \and
 J.~A.~Barrio\inst{2} \and
 D.~Bastieri\inst{7} \and
 J.~Becerra Gonz\'alez\inst{8,}\inst{9}
% ...
}
\institute {IFAE, Edifici Cn., Campus UAB, E-08193 Bellaterra, Spain
 \and Universidad Complutense, E-28040 Madrid, Spain
 \and INAF National Institute for Astrophysics, I-00136 Rome, Italy
 \and Universit\`a  di Siena, and INFN Pisa, I-53100 Siena, Italy
 \and Technische Universit\"at Dortmund, D-44221 Dortmund, Germany
 \and Max-Planck-Institut f\"ur Physik, D-80805 M\"unchen, Germany
 \and Universit\`a di Padova and INFN, I-35131 Padova, Italy
 \and Inst. de Astrof\'{\i}sica de Canarias, E-38200 La Laguna, Tenerife, Spain
 \and Depto. de Astrof\'{\i}sica, Universidad de La Laguna, E-38206 La Laguna, Spain
% ...
}
\offprints{corresponding author D.~Hadasch (hadasch@ieec.uab.es)}

%   \date{Received September 15, 1996; accepted March 16, 1997}

  \abstract
   {The high frequency peaked BL Lac PKS 2155-304 with a redshift of  \textit{z}=0.116 was
discovered in 1997 in the very high energy (VHE, E$>$100\,GeV) $\gamma$-ray range by the University of Durham Mark VI $\gamma$-ray
 Cherenkov telescope in Australia with a flux corresponding to 20\% of the Crab Nebula
flux. It was later observed and detected with high significance by the
Southern Cherenkov observatory H.E.S.S. establishing this source as the
best studied Southern TeV blazar. Detection from the Northern hemisphere is difficult due to challenging observation conditions under large zenith
angles.  In July 2006, the H.E.S.S. collaboration reported an extraordinary
outburst of VHE $\gamma$-emission. During the outburst, the VHE $\gamma$-ray
emission was found to be variable on the time scales of minutes and with a mean
flux of $\sim$7 times the flux observed from the Crab Nebula.
Follow-up observations with the MAGIC-I standalone Cherenkov telescope were triggered by this extraordinary outburst
and \mbox{PKS\,2155-304} was observed between 28 July to 2 August 2006 for 15 hours at large zenith angles.}
   {We studied the behavior of the source after its extraordinary flare. Furthermore, we developed an analysis method in order to analyze these data taken under large zenith angles.}
{Here we present an enhanced analysis method for data taken at high zenith angles. We developed improved methods for event selection that led to a better background suppression.}  
   {The quality of the results presented here is superior to the
results presented previously for this data set: detection of the source on a
higher significance level and a lower analysis threshold. 
The averaged energy spectrum we derived has a spectral index of ($-3.5\pm0.2$)
above 400\,GeV, which is in good agreement  with the spectral shape measured by
H.E.S.S. during the major flare on MJD 53944. Furthermore, we present the
spectral energy distribution modeling of \mbox{PKS\,2155-304}. With our
observations we increased the duty cycle of the source extending the light
curve derived by H.E.S.S. after the outburst. Finally, we find night-by-night
variability with a maximal amplitude of a factor three to four and an
intranight variability in one of the nights (MJD 53945) with a similar
amplitude. }
   {}

   \maketitle
%
%________________________________________________________________

\section{Introduction}

The blazar \mbox{PKS\,2155-304} is the so-called lighthouse of the Southern hemisphere. The
high frequency peaked BL Lac \mbox{PKS\,2155-304}, at a redshift of \textit{z}=0.116, was
discovered in the VHE $\gamma$-ray range by the University of Durham Mark VI $\gamma$-ray
 Cherenkov telescope (Australia) in 1997 with a flux corresponding to $\sim$0.2 times the Crab Nebula
flux \citep{durham}.
\mbox{PKS\,2155-304} was confirmed as a TeV $\gamma$-ray source by the H.E.S.S. group after observations in 2002 and 2003 \citep{hess1}.
In July 2006, the H.E.S.S. collaboration reported an
extraordinary outburst of VHE $\gamma$-emission \citep{hess2155}. During this outburst, the
$\gamma$-ray emission was found to be variable on time scales of minutes with a mean
flux of $\sim$7 times the flux observed from the Crab Nebula for E$>$200\,GeV.
Large amplitude flux variability at
these time scales implies that the TeV emission originates from a small region due to the requirement that light travel times must be sufficiently short in the frame of the emitting region. Follow-up observations
of the outburst by the MAGIC telescope were triggered in a Target of
Opportunity program by an alert from the H.E.S.S. collaboration \citep{atel}. The results of this campaign are presented in this paper.
The CANGAROO group also observed the source
immediately after the flare, obtaining a significance of
4.8\,$\sigma$ and an averaged integral flux above 660\,GeV that corresponds to $\sim$45\% of the flux observed from the Crab Nebula
\citep{saka}.
The H.E.S.S. collaboration continued observations and detected 44
hours later again a major VHE flare. The data were taken
contemporaneously with the Chandra satellite and a strong correlation between the
X-ray and the VHE $\gamma$-ray bands was found \citep{hess_2}.
MAGIC observed on six consecutive nights following the trigger and here we present the final results of the data set.
Due to observational constraints, MAGIC did not observe the source during the major flares, but in two cases data were taken immediately afterwards,
part of the data being simultaneous with H.E.S.S. and Chandra data.
Two years later, in 2008 another multi-wavelength campaign was performed providing simultaneous MeV-TeV data taken by the Fermi Gamma-ray Space Telescope
and the H.E.S.S. experiment \citep{hess_fermi}. 
With these data the low state  of the source could be modeled, including
high energy data for the first time.
All these observations  establish this source
as the best studied Southern TeV blazar. 

Blazars are Active Galactic Nuclei whose relativistic plasma jets nearly point towards the observer. The overall (radio to $\gamma$-ray)
spectral energy distribution (SED) of these objects shows two broad non-thermal continuum peaks. For high energy peaked BL Lac objects (HBLs), the first peak of the SED covers the UV/X-ray bands whereas the second peak is in the multi GeV band.
There are various models to explain this spectral shape. They are generally divided into two classes: leptonic and hadronic.
Both models attribute the peak at keV energies to synchrotron radiation from relativistic electrons (and positrons) within the jet,
but they differ on the origin of the TeV peak. The leptonic models advocate the
inverse Compton scattering mechanism, utilizing synchrotron self Compton (SSC)
interactions and/or inverse Compton interactions with an external photon field,
to explain the VHE emission, (e.g. \citet{maraschi,dermer,sikora}). 
On the other hand, hadronic models
account for the VHE emission
through initial p-p or p-gamma interactions
or via proton synchrotron emission (e.g. \citet{mannheim,aharonian,pohl}).

Blazars often show violent flux variability, which may or may not be correlated between the different energy bands.
Strictly simultaneous observations are crucial to investigate these correlations and understand the underlying physics of blazars.

The structure of the paper is the following: In section~\ref{magic} we introduce the MAGIC telescope. In section~\ref{data} we present a new analysis method optimized for large zenith angle (ZA) observations. We test the method on Crab Nebula data taken under large zenith angles (60$^\circ$ to
66$^\circ$) in section~\ref{crab_1}. In section~\ref{2155_1} we apply the method to the \mbox{PKS\,2155-304} data set and in addition, we model the spectrum in section~\ref{sed}. We summarize the results in section~\ref{conc}.

\section{The MAGIC telescope}\label{magic}

The MAGIC collaboration operates two 17\,m diameter Imaging Cherenkov Telescopes
on the Canary Island of La Palma. The data set presented here was taken in 2006, i.e. before the second MAGIC telescope was installed. Therefore, only single telescope data are available for this analysis.
The camera of the MAGIC phase I telescope has hexagonal shape with a
field of view (FoV) of $\approx$\,3.5$^\circ$ mean diameter and comprises 576
high-sensitivity photomultiplier tubes. 
The energy resolution is
$\Delta$E/E=20\% above 200\,GeV. The single telescope flux sensitivity for a point-like source is 1.6\%
of the Crab Nebula flux for a 5$\sigma$ detection in 50\,hrs of on-source time.
The energy threshold is about 50-60\,GeV at the trigger level. 
Further details of telescope parameters and on the performance can be found in \citet{baixeras,cortina,crab_60}.
These performance values are valid for observations at small zenith angles, where the
distance between the extended air shower and the telescope is the shortest.

In the case of \mbox{PKS\,2155-304} the observations had to be conducted at high ZA (up to 66$^\circ$) since this source culminates at 58$^\circ$ ZA in La Palma.
Under these special conditions a larger effective area is achieved and sources from a large section
of the Southern sky can be observed with a threshold of a few hundred GeV
($\approx$\,100\,GeV-500\,GeV, zenith angle dependent).
Observations at such high ZA not only produce a significantly higher
threshold, but also usually result in a considerable loss of sensitivity. The
analysis presented here is a re-analysis of the data presented in \cite{mazin_proc}.
The goal of
this re-analysis was to improve the results, obtain a more significant
detection and a lower analysis threshold.

%%%%%%%%%%%%%%%%%%%%%%%%%%%%%%%%%%%%%%%%%%%%%%%%%%%%%%%%%%%%%%
%                                                            %
%                                                            %
%                                                            %
%                  D A T A   A N A L Y S I S                 %
%                                                            %
%%%%%%%%%%%%%%%%%%%%%%%%%%%%%%%%%%%%%%%%%%%%%%%%%%%%%%%%%%%%%%

\section[data]{Data Analysis}\label{data}

All data analyzed in this work were taken in wobble mode, i.e. tracking a
sky direction, which is 0.4$^\circ$ off the source position and alternating it
every twenty minutes to the opposite side of the camera center. These changes
avoid effects caused by camera inhomogeneities. The background is
estimated from the mirrored source position in the same FoV, which improves the
background
estimation and yields a better time coverage because no extra OFF data have to
be taken. In this analysis we use three OFF regions which are distributed at angles of 90$^\circ$, 180$^\circ$ and 270$^\circ$ from the source position with respect to the camera center.
These OFF regions have the same size and the same distance from the camera center.

The analysis presented here improves on the original one of
\cite{mazin_proc} in several aspects. It makes use of the pixel-wise timing
information for the image cleaning, as well as in the background suppression,
through the gradient of the signal arrival time along the major image axis.
These improvements allow to reduce the energy threshold for these high zenith
angle observations from $\sim$600\,GeV to $\sim$300\,GeV, and increase the overall significance
of the excess from 11 to 25 standard deviations. In the following the analysis is explained in detail.

In this work we use the time image cleaning \citep{image_clean}: 
given the sub-nsec timing resolution of the data acquisition
system and thanks to the parabolic structure of the telescope mirror
a small integration window can be chosen.
This reduces the number of pixels with signals due to night sky background which
survive the image cleaning. A minimum number of 6 photoelectrons in the core pixels and 3 photoelectrons in the boundary pixels of the images are required. Differences between the signal arrival times have to be smaller than 1.75\,ns. 
This allows a reduction in the pixel threshold level
(i.e. retaining pixels with less charge) of the image cleaning, leading
to a lower analysis energy threshold.  
The cleaned camera image is characterized by a set of image parameters based on \citet{hillas}. These parameters provide a geometrical description of the images of the showers and are
used to infer the energy of the primary particle, its arrival
direction and to distinguish between $\gamma$-ray showers and hadronic showers.

\begin{figure}[h]
    \begin{center}
    \includegraphics*[width=0.5\textwidth,angle=0,clip]{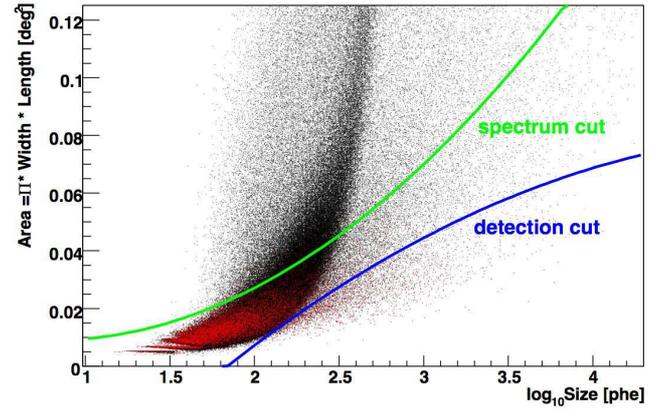}
\caption[Spectrum and detection cuts]{\label{fig:cuts_mc}\textit{Red: }Monte Carlo gammas at high ZA (60$^\circ$-66$^\circ$), \textit{black: }Background events before cuts, \textit{blue: }AREA cut used for the detection ($\theta^{2}$ plot) of the Crab Nebula, \textit{green: }Specialized area parameterization used for the spectrum of the Crab Nebula.} 
    \end{center}
\end{figure}

Cosmic-ray background suppression is achieved by means
of dynamical cuts in \textit{AREA}\footnote{\textit{AREA}=$\pi$$\times$\textit{W}$\times$\textit{L}, \textit{W} and \textit{L} being the width and length of the image as defined in \cite{hillas}.}
versus \textit{SIZE}\footnote{Total number of measured photo electrons in the image. This value is roughly proportional to the energy of the primary particle for a given ZA.} \citep{Riegel}. Typically, events with low \textit{SIZE}
are rejected because the $\gamma$/ hadron separation is very poor for such
images. A standard \textit{AREA} cut for low zenith angle data is shown in
Figure \ref{fig:cuts_mc} by the blue line. This standard cut removes
low \textit{SIZE} events corresponding to events with low energies. 

The angular distance, $\theta$, between
the reconstructed direction of an event and the catalog position of the 
 $\gamma$-ray source is an
essential background rejection parameter.
The event direction is obtained by using the DISP parameter
\citep{lessard,domingo}, which uses the image shape to estimate where,
along the major axis of the image, lies the point in the
camera that corresponds to the shower direction.
For a point-like $\gamma$-ray source, the distribution of $\theta^2$ will peak at around zero, whereas background events produce a rather flat distribution.

Analysis of data taken at high ZA requires special treatment compared to low ZA data.
The Cherenkov light of the showers observed at large ZA has a longer optical path, as it has to pass a thicker layer of atmosphere.
Therefore, the shower maximum is located farther from the observatory and
the photon density of each shower decreases. This reduces the number of detectable
showers, especially at low energies. The global effect is a shift of the energy
threshold to higher energies with increasing ZA.
At the same time, images taken at high ZA correspond, for a given value of \textit{SIZE}, to events with higher amount of shower particles than at low ZA.
The higher amount of shower particles reduces the intrinsic shower fluctuations.
For this reason one obtains better defined $\gamma$-ray showers at high ZA. 
This implies that the minimum \textit{SIZE} at which effective background suppression is possible can be lowered for high ZA compared to low ZA.
In this section we quantify this effect.

A specialized parameterization of the \textit{AREA} cut was developed for this high ZA study.
Together with improving the cosmic-ray background rejection power,
its main objective was to lower the energy threshold.
A comparison between the  standard \textit{AREA} cut, which is used for the detection of the source, and the specialized parameterization, used for obtaining the spectrum, is shown in Figure
\ref{fig:cuts_mc}. 
For a low energy analysis the parameterization of the spectrum cut is:
\begin{equation}
\textit{Area}<(1.4\cdot (\log_{10}(\textit{SIZE})-0.8)^{2}+1)\cdot x ,
\end{equation} 
where x is varied between 0.007 and 0.009 to study the dependency of the
spectrum from the cut efficiencies. The red
points in Figure~1 represent Monte Carlo (MC) gammas and the black ones background events before cuts.
The standard \textit{AREA} cut used for the detection and the spectrum
\textit{AREA} cut are shown by the blue and the green parabolas, respectively.
For the spectrum \textit{AREA} cut, it can been seen that events with small
\textit{SIZEs} survive the background suppression.

A quantitative estimate for the background suppression power can be made with the quality factor \textit{Q}:
\begin{equation}
Q=\frac{\epsilon_{\gamma}}{\sqrt{\epsilon_{bg}}} ,
\end{equation} 
where $\epsilon_{\gamma}$ is the fraction of $\gamma$-rays that are retained after a cut. $\epsilon_{bg}$ is the corresponding fraction of retained background events.
To check that the background suppression with a combination of the new
\textit{AREA} cut and a $\theta^{2}$ cut is working for small
\textit{SIZE}-values, we calculate the \textit{Q}-factor for the four lowest
bins in \textit{SIZE} (see table~\ref{table:qfactor}). For the \textit{AREA}
cut we require a $\gamma$-efficiency of 0.9 and for the $\theta^{2}$ cut one
of 0.5.
Having \textit{Q}-factors (QHZA) in excess of 1 in each bin
implies that the applied cut works efficiently.
Comparing Q$_{HZA}$ with the one obtained for low ZA (Q$_{LZA}$) shows a stronger background suppression power for the high ZA case.
We also look closer into the rejection power of the \textit{AREA} and $\theta^{2}$ cuts individually.
First, we compare the hadron-efficiencies for the \textit{AREA} cut at high ($\epsilon^{LZA}_h$) and low ZA ($\epsilon^{HZA}_h$).
There is no difference between the two for
$\log_{10}(\textit{SIZE}) < 2.0$ and there is a clear advantage for a high ZA analysis at
higher SIZE values.
Next we compare the $\theta^{2}$-distributions of MC
$\gamma$-events for the lowest \textit{SIZE}-bins at high and low ZA.
We calculate the ratio of the $\theta^{2}$ cut values with 50\%  $\gamma$-ray event efficiency ($\theta^{2}_{LZA,50\%}$/$\theta^{2}_{HZA,50\%}$, see table~\ref{table:qfactor}).
This fraction is directly proportional to the improvement of the background rejection at high ZA with respect to low ZA since the background has a flat distribution in the $\theta^{2}$-plot at $\theta^{2} < 1$\,deg$^2$.
We obtain roughly 2 times better background suppression at high ZA than at small ZA for small \textit{SIZE} values.

\begin{table}[ht]
\begin{center}
\caption{\label{table:qfactor}Quality factors for high and low ZA analyses derived by using MC $\gamma$-events and real background events. \textit{Q}-factor at high ZA: Q$_{HZA}$; \textit{Q}-factor at low ZA: Q$_{LZA}$; hadron-efficiency of the \textit{AREA} cut at low/ high ZA  ($\gamma$-efficiency = 0.9): $\epsilon^{LZA}_h/\epsilon^{HZA}_h$, $\theta^{2}$-values at low/ high ZA ($\gamma$-efficiency = 0.5): $\theta^{2}_{LZA,50\%}$/$\theta^{2}_{HZA,50\%}$}
\begin{tabular}{c c c c c}
\hline
\textit{SIZE}-bin               & Q$_{HZA}$ &  Q$_{HZA}$/Q$_{LZA}$  & $\epsilon^{LZA}_h/\epsilon^{HZA}_h$   & $\theta^{2}_{LZA,50\%}$/$\theta^{2}_{HZA,50\%}$ \\ 
     $\log_{10}(\textit{SIZE})$ &           &                       &                                       &                                                 \\ \hline \hline
1.5 -- 1.75                     & 1.06      &  1.25                 & 1.0                                   & 1.92                                            \\
1.75 -- 2.0                     & 1.51      &  1.45                 & 1.0                                   & 2.30                                            \\
2.0 -- 2.25                     & 2.27      &  1.49                 & 1.13                                  & 2.58                                            \\
2.25 -- 2.5                     & 4.44      &  1.73                 & 2.0                                   & 2.0                                            \\
\hline
\end{tabular}
\end{center}
\end{table}

In addition to the \textit{AREA} cut, cuts are applied to timing parameters,
which describe the time evolution along the major image
axis and the RMS of the time spread.
These two additional parameters
lead to better background suppression,
yielding a better sensitivity \citep{image_clean}. 

A possible $\gamma$-ray signal coming from point-like sources can be identified
with the directional information of the parameter $\theta$.
The necessary signature of a $\gamma$-ray signal is
an excess at small $\theta^2$ values, usually lower than 0.04\,deg${^2}$.
The cuts used to calculate the significance of the detection and the cuts used to derive the energy spectrum of \mbox{PKS\,2155-304}
were optimized on a Crab Nebula sample taken at large zenith angles (see~\ref{crab_1} for details).

The primary $\gamma$-ray energies were reconstructed from the image parameters using a Random Forest regression method \citep{random} trained with MC simulated events \citep{knapp,majum}.
The MC sample is characterized by a power-law spectrum between 10\,GeV and 30\,TeV with a differential spectral photon index of $\alpha$=-2.6. The events were selected to cover the same ZA range as the data.
Compared to the previous analysis \citep{mazin_proc} improvements are obtained
because of an updated MC sample at high zenith angles 
leading to better agreement between data and MC.

Since the analysis technique described in this section is new, we first test its performance on a data set 
from a known, bright and stable $\gamma$-ray emitter: the Crab Nebula 
before applying the method to \mbox{PKS\,2155-304}.
 
%%%%%%%%%%%%%%%%%%%%%%%%%%%%%%%%%%%%%%%%%%%%%%%%%%%%%%%%%%%%%%
%                                                            %
%                                                            %
%                  R E S U L T S                             %
%                                                            %
%                                                            %
%%%%%%%%%%%%%%%%%%%%%%%%%%%%%%%%%%%%%%%%%%%%%%%%%%%%%%%%%%%%%%

\section{Results}
\subsection{Crab Nebula}\label{crab_1}

The Crab Nebula is one of the best studied celestial objects because of the
strong persistent emission of the Nebula over 21 decades of frequencies. 
It was the first object to be detected
at TeV energies by the Whipple collaboration in 1989
\citep{weekes_crab} and is the strongest steady source of VHE $\gamma$-rays. Due to the
stability and the strength of the $\gamma$-ray emission the Crab Nebula is
generally considered the standard candle of the TeV $\gamma$-ray astronomy. The
measured $\gamma$-ray spectrum extends from 60\,GeV \citep{crab_60} up to 80\,TeV
\citep{crab_80} and 
appears to have maintained a constant flux
in the VHE range over the years (from 1990 to present).

\subsubsection{Data set and analysis sensitivity}
\begin{figure}[ht]
  \begin{center}
    \includegraphics*[width=0.49\textwidth,angle=0,clip]{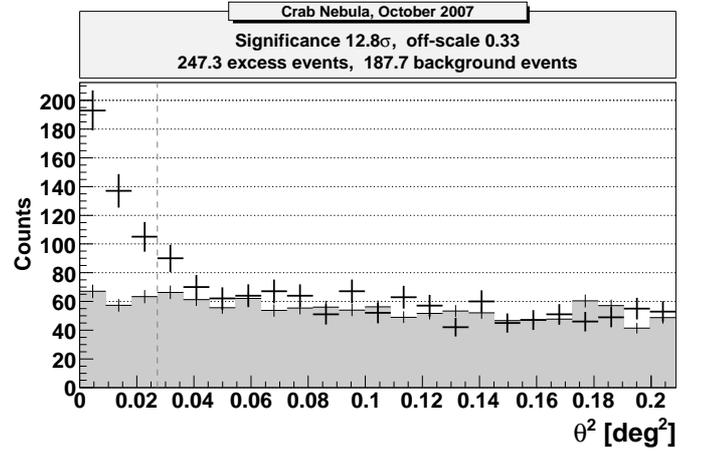}
    \caption{\label{fig:crab_theta}The ON-source and normalized background distribution of $\theta^{2}$. The ON-source is shown in the black crosses and the background is shown in the gray shaded region. 2.15\,hrs of Crab data show an excess with a significance of 12.8$\sigma$.}
  \end{center}
\end{figure}

In October 2007, the MAGIC telescope took Crab Nebula data with a zenith
angle range of 60$^\circ$ up to 66$^\circ$. The data were taken under dark sky conditions and in
wobble mode. After quality cuts an effective on-time of 2.15\,hrs is obtained.
Using detection cuts presented in section~\ref{data}, Fig.~\ref{fig:cuts_mc} (i.e. optimized on significance of a separate Crab Nebula
sample also taken at high zenith angles), a total of 247 excess events above 187 background events have been detected. The number of background events is
obtained from three equally sized OFF regions
and applying a geometrical scale factor of 0.33 (see Fig.~\ref{fig:crab_theta}). 
The significance of this $\gamma$-ray signal,
obtained using Eq. 17 of \citet{lima}, is 12.8\,$\sigma$. This corresponds to an
analysis sensitivity of 8.7 $\frac{\sigma}{\sqrt{hour}}$. Using the same set of cuts we obtain the
following sensitivities for integral fluxes ($\Phi$):
\begin{eqnarray*}
\Phi(E>0.4\,\mathrm{TeV}) &\Rightarrow& 5.7\% \mathrm{\,\,Crab\,\,in\,\,50\,hrs} \\
\Phi(E>0.63\,\mathrm{TeV}) &\Rightarrow& 5.6\% \mathrm{\,\,Crab\,\,in\,\,50\,hrs} \\
\Phi(E>1.0\,\mathrm{TeV}) &\Rightarrow& 5.9\% \mathrm{\,\,Crab\,\,in\,\,50\,hrs} \\
\Phi(E>1.5\,\mathrm{TeV}) &\Rightarrow& 6.8\% \mathrm{\,\,Crab\,\,in\,\,50\,hrs}
\end{eqnarray*}

\subsubsection{Differential Energy Spectrum}
\begin{figure}[ht]
  \begin{center}
    \includegraphics*[width=0.49\textwidth,angle=0,clip]{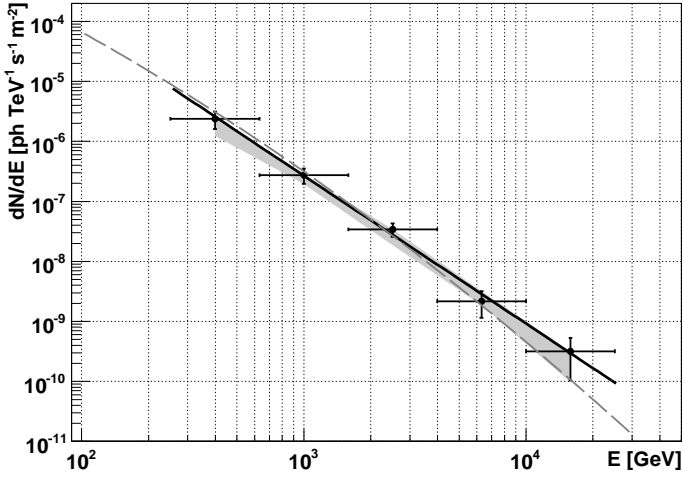}
    \caption{\label{fig:crab_spec}Differential energy spectrum of the Crab Nebula. Black line: power law fit to the data, gray band: systematic uncertainties of the analysis, dashed line: published data taken at low zenith angles \citep{crab_60}.}
  \end{center}
\end{figure}

Using the data set described in the previous section, we derived the differential energy spectrum of the Crab
Nebula (see Fig.~\ref{fig:crab_spec}). The spectrum is well described by a
simple power law of the form:
\begin{small}
\begin{eqnarray}
\frac{\mathrm{d}N}{\mathrm{d}E}=(2.7\pm0.4)\cdot10^{-7}\left(\frac{\mathrm{E}}{\mathrm{TeV}}\right)^{-2.46\pm0.13}\left(\frac{\mathrm{ph}}{\mathrm{TeV}\,\mathrm{s}\,\mathrm{m}^{2}}\right). \nonumber
\end{eqnarray}
\end{small}
The errors are statistical only.
The gray band represents the
range of results obtained by varying the total cut efficiency between 40\% and
70\%, i. e. the cut efficiency after applying the \textit{AREA} and the $\theta^{2}$ cuts. For comparison, the Crab Nebula spectrum from
data taken at low zenith angles is drawn as a dashed line \citep{crab_60}. A very good
agreement has been found.  

\subsection{\mbox{PKS\,2155-304}}\label{2155_1}

\begin{figure}[ht]
  \begin{center}
    \includegraphics*[width=0.49\textwidth,angle=0,clip]{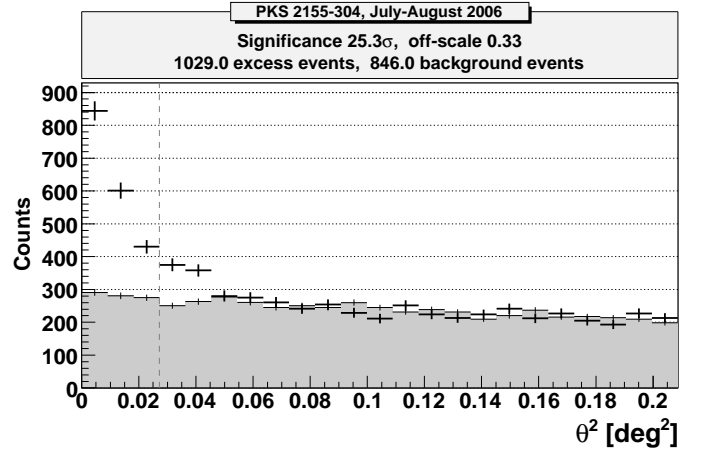}
    \caption{\label{fig:2155_theta}The ON-source and normalized background distribution of $\theta^{2}$. The denotations are the same as in Figure \ref{fig:crab_theta}. A clear excess with a significance of more than 25 standard deviations for a source at the position of \mbox{PKS\,2155-304} is found.}
  \end{center}
\end{figure}

The MAGIC telescope observed the blazar \mbox{PKS\,2155-304} from 28 July to 2 August
2006 (MJD 53944.09 -- MJD 53949.22) over
the zenith angle range of 59$^\circ$ to 64$^\circ$. The data were taken under dark sky
conditions and in wobble mode. After quality cuts a total effective on-time of
8.7\,hrs is obtained. For the detection of \mbox{PKS\,2155-304}, the same cuts are used as
for the detection of the Crab Nebula. Three OFF regions are used and 1029
excess events above 846 normalized background events are detected. A significance of 25.3
standard deviations is obtained, whereas with the previous analysis only 11$\sigma$ could be achieved \footnote{The earlier analysis in \cite{mazin_proc} made use of Random Forest for background suppression, but using only shape image parameters; the analysis presented here uses also time parameters, which later became standard \citep{image_clean}.}. The corresponding $\theta^{2}$-plot is presented in Fig.~\ref{fig:2155_theta}.

\subsubsection{Differential Energy Spectrum}
The differential energy spectrum for the whole data set is shown in Fig. 5 as a black line 
together with the spectrum of H.E.S.S. (dashed line),
measured during the strong outburst \citep{hess2155}. Note that H.E.S.S. and MAGIC data are not simultaneous. The
spectral points obtained in this analysis are fitted in the energy range from 400\,GeV to 4\,TeV, because at
lower energies H.E.S.S. reported a change of the slope ($-3.53\pm0.05$ above
400\,GeV to $-2.7\pm0.06$ below 400\,GeV). The fitted MAGIC data points are consistent
with a power law:
\begin{small}
\begin{eqnarray}
\frac{\mathrm{d}N}{\mathrm{d}E}=(1.8\pm0.2)\cdot10^{-7}\left(\frac{\mathrm{E}}{\mathrm{TeV}}\right)^{-3.5\pm0.2}\left(\frac{\mathrm{ph}}{\mathrm{TeV}\,\mathrm{s}\,\mathrm{m}^{2}}\right) \nonumber
\end{eqnarray}
\end{small}
with a fit probability after the $\chi^{2}$-test of 81\%.  Above 400\,GeV, the
energy flux measured by H.E.S.S. from the preceding flare of \mbox{PKS\,2155-304}
is one order of magnitude higher than the flux measured by MAGIC.
It is interesting to note that although the flux measured with H.E.S.S. is higher, the
spectral slope remains the same within the
statistical errors.

\begin{figure}[ht]
  \begin{center}
    \includegraphics*[width=0.49\textwidth,angle=0,clip]{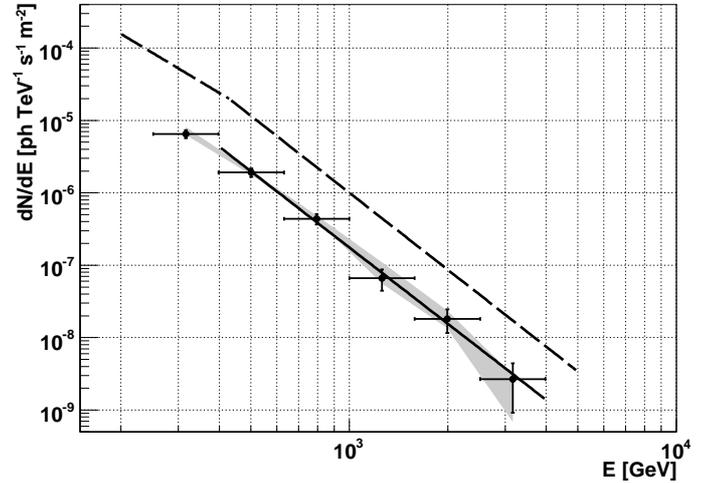}
    \caption{Differential energy spectrum (black line) for the whole data set together with
systematic errors obtained by varying cuts efficiencies (gray band). The black
dashed line corresponds to the H.E.S.S. measurement during the flare.}
  \end{center}
\end{figure}

\begin{figure}[ht]
  \begin{center}
    \includegraphics*[width=0.49\textwidth,angle=0,clip]{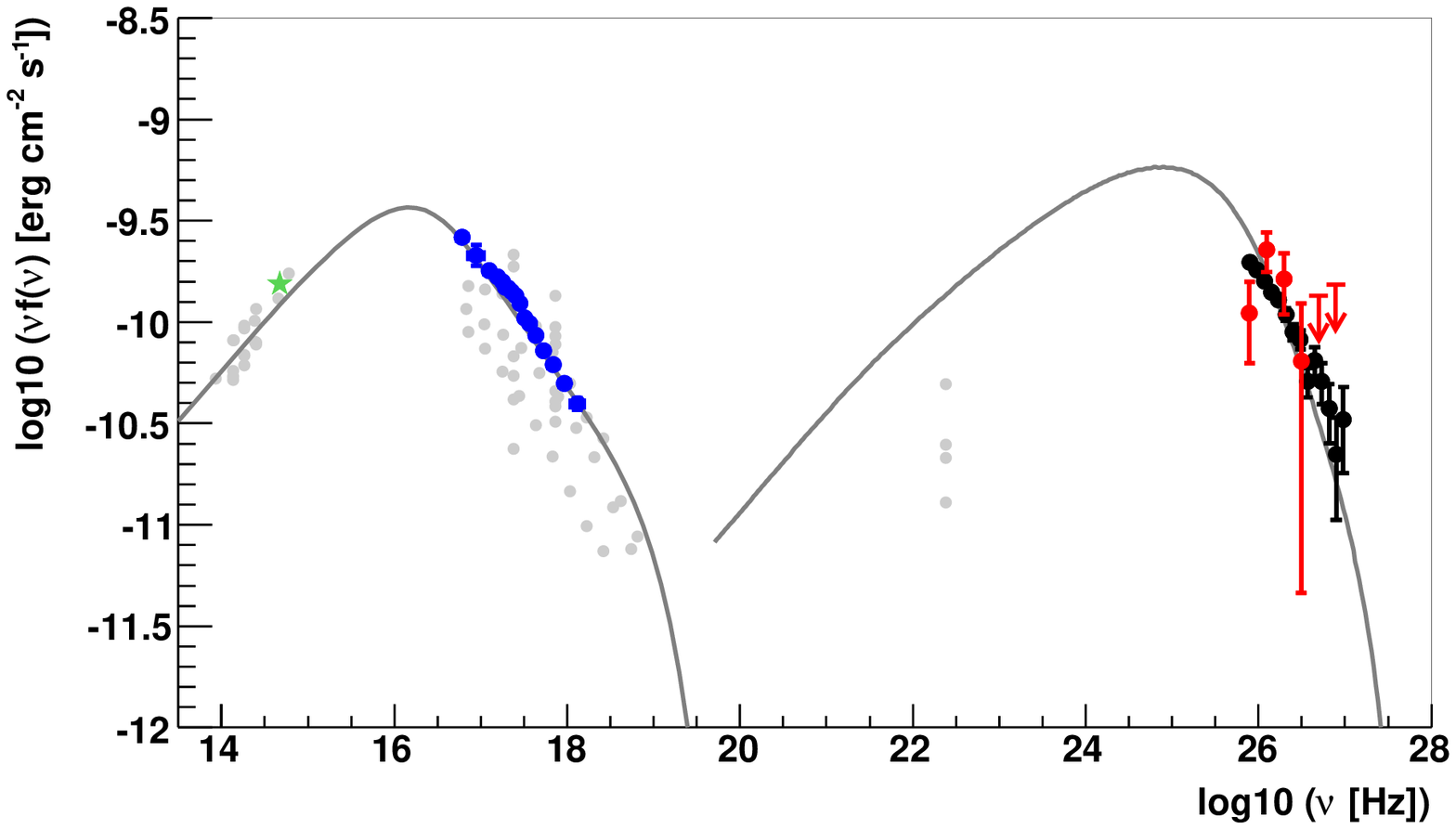}
\caption{\label{fig:sed}The overall spectral energy distribution (SED) of \mbox{PKS\,2155-304} from optical wavelengths through VHE $\gamma$-rays. The gray line denotes an SSC model as in \cite{kraw}. The effect of the EBL is taken into account by using the model of \cite{franc}. Red points: MAGIC, red arrows: MAGIC 2\,$\sigma$ upper limits, black: H.E.S.S., blue: Chandra, green star: optical point (ROTSE). All data were taken simultaneously on the day MJD 53946 with exception of the ROTSE point and they are taken from \cite{hess_2}. Selected historical data from infrared to $\gamma$-rays are shown in gray for comparison reasons \citep{beppo}.} 
\end{center} 
\end{figure}

\subsubsection{Account of EBL attenuation in $\gamma$-ray spectra}\label{ebl}

For the VHE-spectra in Fig.~\ref{fig:sed}, representing data from the day MJD 53946, the absorption effect caused by the extragalactic background light (EBL) was taken into account.
The VHE photons 
interact with the low-energy photons of the EBL
\citep{gould, hauser}. The predominant reaction
$\gamma_{VHE}+\gamma_{EBL}\rightarrow e^{+}e^{-}$ leads to an attenuation of
the intrinsic AGN spectrum $\mathrm{d}N/\mathrm{d}E_{intr}$ that can be described
by
\begin{small}
\begin{eqnarray}
\mathrm{d}N/\mathrm{d}E_{obs}=\mathrm{d}N/\mathrm{d}E_{intr} \cdot \exp[-\tau_{\gamma \gamma}(E, z)] \nonumber
\end{eqnarray}
\end{small}
with the observed spectrum $\mathrm{d}N/\mathrm{d}E_{obs}$, and the energy
dependent optical depth $\tau_{\gamma \gamma}(E, z)$.
We apply the EBL model of \citet{franc} which is the same as the H.E.S.S. collaboration used to account for EBL attenuation in their spectrum, and which agrees well with other state-of-the-art EBL models
\citep{2010A&A...515A..19K,2011MNRAS.410.2556D,2011arXiv1104.0671G}.

\subsubsection{Spectral energy distribution and SSC modeling}\label{sed} 

SSC models have been very successful in describing the observed HBL
multi-frequency spectra. In the homogeneous one-zone SSC model, the X-ray
emission comes from synchrotron radiation emitted by a population of high
energy electrons, followed by inverse Compton scattering of synchrotron photons
to TeV energies, which explains the $\gamma$-ray emission from TeV blazars.
Based on this model it is possible to constrain the parameter space of the
emission region and estimate its basic parameters, the Doppler factor, $D$, and
the rest-frame magnetic field, $B$, of the emitting plasma in the relativistic
jet.

The acceleration region is approximated by a spherical "blob" with radius $R$
and bulk Lorentz factor $\Gamma$ that is moving along the jet under a small
angle $\theta$ to the line of sight.
The blob contains relativistic electrons of density $\rho$ accelerated by shock
acceleration processes. It is assumed that the energy spectrum of the electrons in
the jet frame can be described by a broken power law with low-energy ($E_{min}$
to $E_{break}$) and high-energy ($E_{break}$ to $E_{max}$) indices $n_{1}$
and $n_{2}$, respectively ($n_{i}$ is from $\mathrm{d}N/\mathrm{d}E\propto
E^{-n_{i}}$; E is the electron energy in the jet frame). The electron
spectrum has exponential cut-offs at energies $E_{min}$ and $E_{max}$.

For a successful modeling simultaneous multi-wavelength information is
required. In Fig.~\ref{fig:sed}
almost simultaneous data taken
on 30 July 2006 with MAGIC, H.E.S.S. and Chandra are shown.
The data taken by Chandra are contemporaneous
with those taken with MAGIC.
Please note that we scaled the spectrum obtained by H.E.S.S., since the data are not entirely contemporaneous with the MAGIC data.
The data for the spectrum computed by H.E.S.S. are from MJD 53946.013--53946.129, whereas the MAGIC data are obtained slightly later between MJD 53946.092--53946.186. These time spans are represented in Fig.~\ref{fig:lc_2nd}. 
This difference  is large enough to produce a difference in the
average fluxes by almost a factor of 3.
We scaled the H.E.S.S. spectrum down by this factor and get a good agreement with the MAGIC result.
For the modeling we did not take into account the optical data point from
ROTSE, because this measurement was taken before the high state of the source. 

We modeled the multi-wavelength spectrum (Fig.~\ref{fig:sed}, gray line) with a one-zone, time independent SSC code
from \cite{kraw}. 
This "by-eye" adjustment of model parameters, instead of e.g. $\chi^{2}$-minimization,
is a common procedure with data of this kind
because of degeneracy of the model
and rather large uncertainty in the data.
In Fig.~\ref{fig:lc_2nd} we show in colored bars the time span of data we used for the SED modeling.
Note that the data set we modeled was taken few hours after the flare and no significant variability is seen in H.E.S.S. or MAGIC
data (see Fig.~\ref{fig:lc_2nd}), which justifies usage of a time independent model.  
Still, the time independent model should be taken with caution.
The resulting model parameters shown in Table~\ref{table:sed_param} 
provide a possible solution in the SSC model parameter space.
The parameters are similar to the ones typically obtained for
HBLs, see e.g. \citet{hesssed,mrk2012}: the Doppler factor is high and the magnetic field strength is low. It has
been pointed out that such high Doppler factors are in conflict with what is
seen on VLBA \citep{piner2010}, but this issue is beyond the scope of this
paper.

\begin{table*}[ht]
\begin{center}
\caption{\label{table:sed_param}Parameters for SED modeling}
\begin{tabular}{c c c c c c c c c}
\hline
Doppler factor & $B$ & $R$            & $\rho$                       &  $E_{min}$          & $E_{max}$          & $E_{break}$          & $n1$ & $n2$ \\ 
               & [T] & $[\mathrm{m}]$ & $[\mathrm{particle/cm^{3}}]$ & $\lg (E_{min}[\mathrm{eV}])$ & $\lg(E_{max}[\mathrm{eV}])$ & $\lg(E_{break}[\mathrm{eV}])$ &     &  \\ \hline \hline
50             & $0.085 \times 10^{-4}$ & $0.9 \times 10^{14}$ & 0.07 & 6.3 & 11.5 & 10.2 & 2 & 4 \\
\hline
\end{tabular}
\end{center}
\end{table*}

\begin{figure*}[h]
  \begin{center}
    \includegraphics*[width=0.9\textwidth,angle=0,clip]{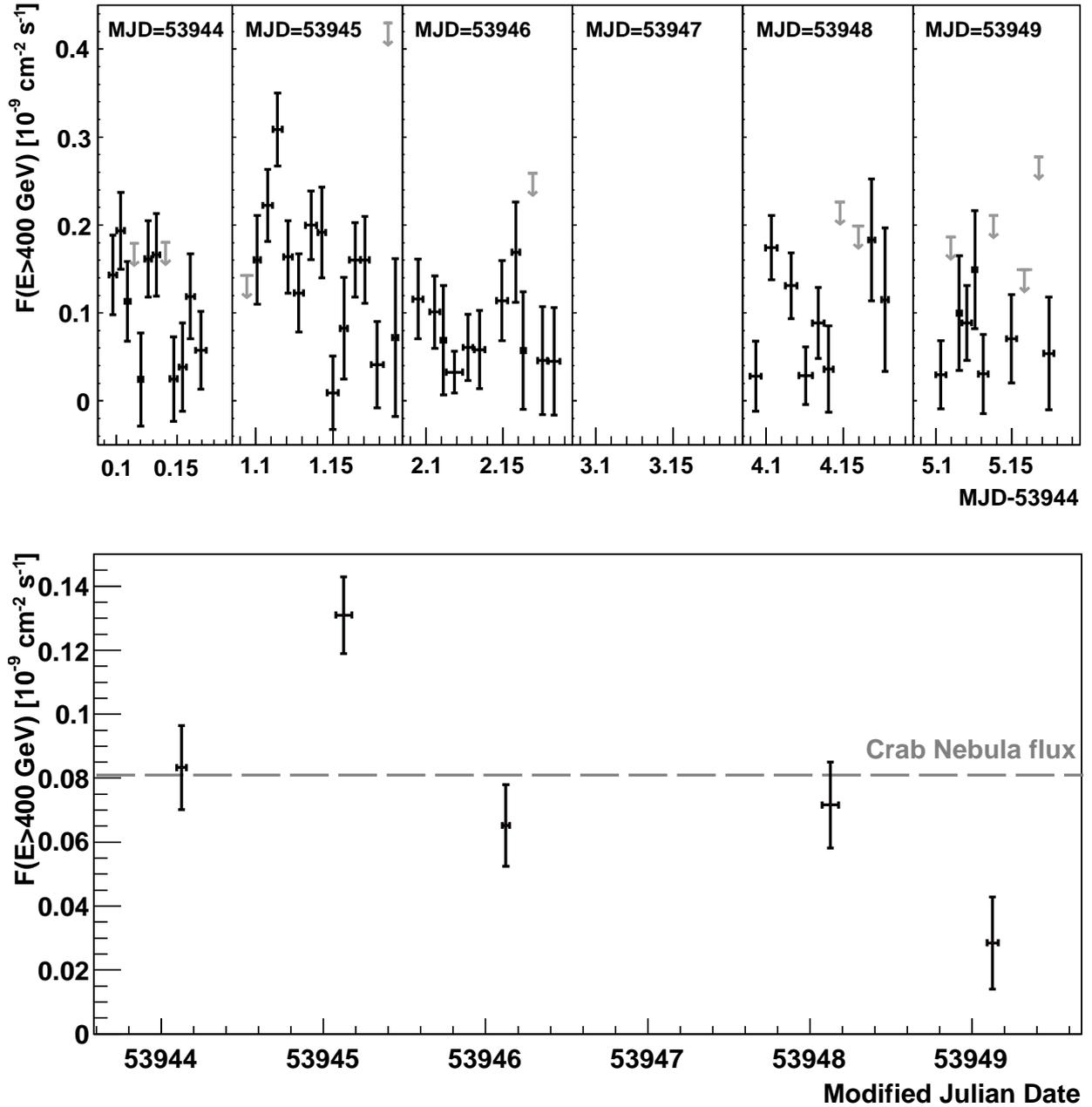}
    \caption{\label{fig:lc}\textit{Top:} MAGIC light curve for individual nights for E$>$400\,GeV of \mbox{PKS\,2155-304}. Only the second
night (MJD\,53945) shows significant intra-night variability. Vertical arrows represent flux
upper limits at a confidence level of 95\%. \textit{Bottom:} Light curve for the total data set for E$>$400\,GeV with one
flux point per night. The error bars in x-direction represent the observation
time. The Crab Nebula flux is shown for comparison. For more information see the text.}
 \end{center}
\end{figure*}

\begin{figure*}[h]
  \begin{center}
    \includegraphics*[width=0.9\textwidth,angle=0,clip]{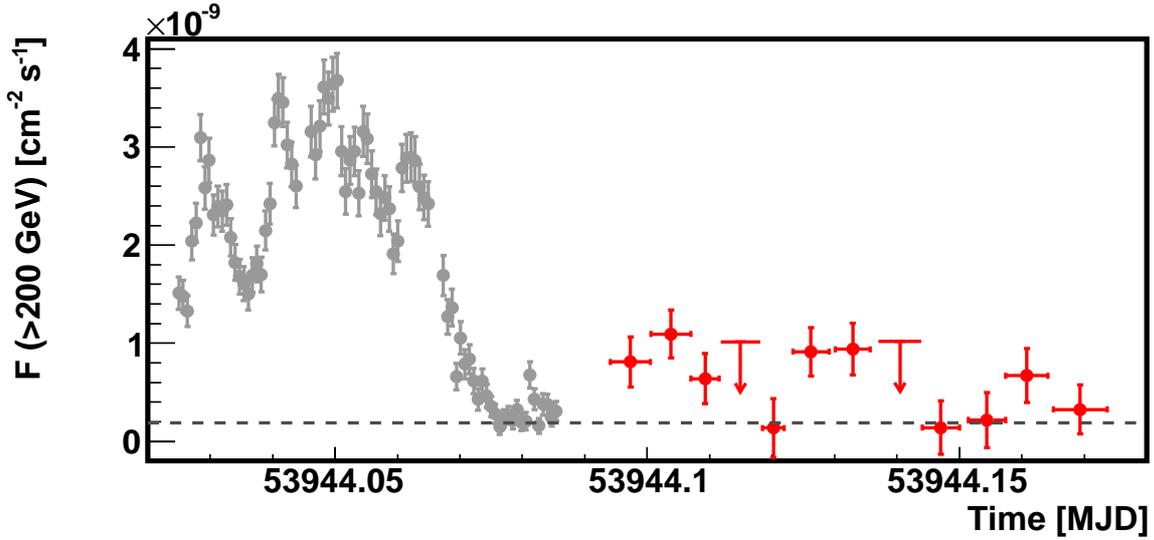}
    \caption{\label{fig:lc_1st}Integral flux above 200\,GeV of the first flare observed from \mbox{PKS\,2155-304} on MJD 53944 versus time measured by H.E.S.S. (gray points) and by MAGIC (red points). The MAGIC points are obtained using the integral flux above 400\,GeV and extrapolated down to 200\,GeV using the derived photon spectral index of -3.5. Instead, assuming the spectral index of -2.7 below 400\,GeV as found by H.E.S.S., the MAGIC points would be 24\% lower in flux. The horizontal line represents the observed flux of the Crab Nebula~\citep{crab_hess}.}
 \end{center}
\end{figure*}

\begin{figure*}[h]
  \begin{center}
    \includegraphics*[width=0.9\textwidth,angle=0,clip]{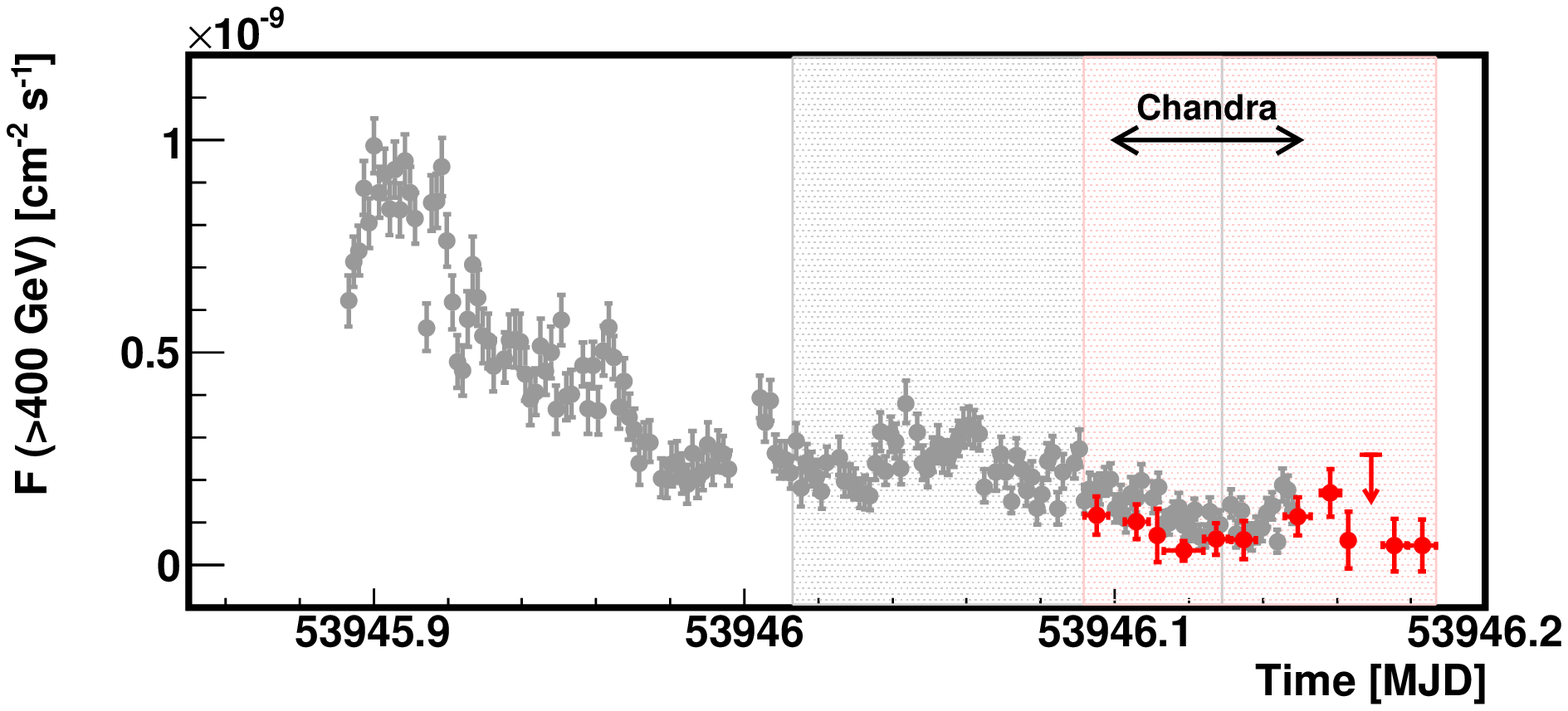}
    \caption{\label{fig:lc_2nd}Light curve of \mbox{PKS\,2155-304} of the second flare observed by MAGIC and H.E.S.S. in the night of July 29-30, 2006. The flux is shown above an energy of 400\,GeV. 
The MAGIC data (red points) are binned in 10 minute intervals
while the H.E.S.S. data (gray points) use four minute intervals.
The red and the gray area represent the time spans when the MAGIC and the H.E.S.S. spectra shown in Fig.~\ref{fig:sed} are derived, respectively. The black arrow shows the time when Chandra observed the source.}
 \end{center}
\end{figure*}

\subsubsection{Light curves}
Figure~\ref{fig:lc} shows the integral light curves for energies above 400\,GeV.
In the upper panel, each point corresponds to an average flux every two data runs (roughly 10 min exposure),
whereas in the lower panel each point corresponds to an average flux per night.
Significant detections in most of the time bins are obtained.
A significant intra-night variability is found for the second night MJD\,53945 (29
July 2006) giving a probability for a constant flux of less than $5\times10^{-9}$. For the
other nights, no significant intra-night variability is found. In the lower panel of Fig.~\ref{fig:lc},
a night-by-night light curve is shown. 
A fit by a constant to the run-by-run light curve results in a chance probability of less than $10^{-12}$. 
However, a fit by a constant to the night-by-night light curve results in a chance probability of $1\times10^{-6}$.
We, therefore, conclude that there is a significant variability on the time scales reaching from days (largest scale we probed)
down to 20 minutes (shortest scale we probed).

In Fig.~\ref{fig:lc_1st} we show the MAGIC light curve with data taken directly
after the first major flare detected by H.E.S.S. (28 July 2006)~\citep{hess2155}. The
MAGIC points are binned in 10 minute intervals
and the H.E.S.S. points have a binning
of one minute intervals \citep{hess_2}. After the large flare, with a measured flux above 200\,GeV of up to 15
times the flux of the Crab Nebula~\citep{hess2155}, the source returned to a lower state, with fluxes of the order of 1 Crab. The second flare observed by
H.E.S.S. is shown in Fig.~\ref{fig:lc_2nd}. MAGIC observed the source in the
low state simultaneous with H.E.S.S. on 30 July 2006. The measurements are in
good agreement concerning the trend of the data points,
but the MAGIC data points appear to lie systematically below the points obtained by H.E.S.S.
The observed difference in the flux is compatible with the systematic uncertainty of the analysis,
which has been estimated to be 20\% of the energy scale or 50\% on the flux level.
Typically, the systematic energy scale uncertainty for analysis of Imaging Cherenkov telescope data 
is at the level of 15\% to 20\%, see e.g. \citep{horns2010}.
The difference we find in this analysis is at the upper edge of the usual systematic uncertainty, which may be due to additional
 systematics of measurements at high zenith
angles where details of the atmosphere are less certain.

%%%%%%%%%%%%%%%%%%%%%%%%%%%%%%%%%%%%%%%%%%%%%%%%%%%%%%%%%%%%%%
%                                                            %
%                                                            %
%                  D I S C U S S I O N                       %
%                                                            %
%                                                            %
%%%%%%%%%%%%%%%%%%%%%%%%%%%%%%%%%%%%%%%%%%%%%%%%%%%%%%%%%%%%%%

\section{Conclusions}\label{conc}

A study of the high-zenith angle performance of the MAGIC
telescope (operating in single-telescope mode) was carried out
with observations of the source \mbox{PKS\,2155-304} during a high state,
conducted between ZA 60$^\circ$ - 66$^\circ$. A new analysis procedure was used in
this
work in order to enhance the sensitivity of
the observations under these special conditions,
for instance, included information about the signal arrival
times to improve the image cleaning and background subtraction.

We tested this new analysis method on a Crab data sample and obtained a sensitivity of 5.7\% of the Crab Nebula flux
for 50\,hrs of observations at high ZA above 0.4\,TeV.
The differential energy spectrum of the Crab Nebula is
in excellent agreement with the published data at lower zenith angles.
This improved analysis is used to reanalyze data of \mbox{PKS\,2155-304} taken with MAGIC in 2006.

The energy spectrum of the whole data set from 400\,GeV up to 4\,TeV has a spectral index of
($-3.5\pm0.2$). As it agrees with the index derived by the H.E.S.S. collaboration during the flaring state of the source, 
we conclude that the spectrum 
does not show any change in its spectral slope with flux state above 400\,GeV.
Furthermore, we corrected the measured spectrum for the effect of the EBL
absorption using the model of \citet{franc} and made an SED modeling of simultaneous data in the VHE and X-ray energy range.

The light curves derived with MAGIC show a significant variability on daily as well as on intra-night time scales.
The MAGIC observations immediately after the extraordinary flare measured by H.E.S.S. on MJD 53944 indicate that the source remained essentially constant for the rest of that night.
The measurements of the MAGIC and the H.E.S.S. experiments are generally in good agreement.

Finally, we conclude that high zenith angle observations with the MAGIC telescope have proven to yield high quality spectra and light curves above 300\,GeV.
With these observations we could extend the duty cycle of \mbox{PKS\,2155-304} observations, 
which is clearly convenient for the study of any flaring source. High-zenith angle observations,
although challenging, generally allow for more uninterrupted coverage of highly variable sources.
Also, future high ZA observations of some objects allow unique spectral 
measurements at higher
energies than is possible at lower ZA
by virtue of the larger effective area.

\begin{acknowledgements}
We would like to thank the Instituto de Astrof\'{\i}sica de
Canarias for the excellent working conditions at the
Observatorio del Roque de los Muchachos in La Palma.
The support of the German BMBF and MPG, the Italian INFN, 
the Swiss National Fund SNF, and the Spanish MICINN is 
gratefully acknowledged. This work was also supported by 
the Marie Curie program, by the CPAN CSD2007-00042 and MultiDark
CSD2009-00064 projects of the Spanish Consolider-Ingenio 2010
programme, by grant DO02-353 of the Bulgarian NSF, by grant 127740 of 
the Academy of Finland, by the YIP of the Helmholtz Gemeinschaft, 
by the DFG Cluster of Excellence ``Origin and Structure of the 
Universe'', by the DFG Collaborative Research Centers SFB823/C4 and SFB876/C3,
and by the Polish MNiSzW grant 745/N-HESS-MAGIC/2010/0.
We thank R. B\"uhler, L. Costamante and B. Giebels for providing H.E.S.S. and multi-wavelength data.
We also thank the anonymous referee for useful comments which helped to improve the manuscript.
\end{acknowledgements}

\bibliographystyle{aa}
\bibliography{/Users/hadasch/Paper/MyPaper/PKS2155/ApJpaper/ver_submitted/bibtex/bibtex_pks2155_2011}

\end{document}